# Scalable Generalized Meta-Spanners Enabling Parallel Multitasking Optical Manipulation

Tianyue Li[1,2,†], Wenyu Gao[3,4,†], Boyan Fu[1,†], Tianhua Shao[1], Yuchao Fu[5], Siarhei Zavatski[5], Jeeban Kumar Nayak[5], Shaohui Yan[3,4], Xiaohao Xu[3,5,*], Shuming Wang[1,*], Baoli Yao[3,4,*], Zhenlin Wang[1], Shining Zhu[1], Olivier J. F. Martin[5], C. T. Chan[2,6,*]

1. National Laboratory of Solid-State Microstructures, School of Physics, Nanjing University, Nanjing, 210093, China.

2. Department of Physics, The Hong Kong University of Science and Technology, Hong Kong SAR, 999077, China.

3. State Key Laboratory of Ultrafast Optical Science and Technology, Xi'an Institute of Optics and Precision Mechanics, Chinese Academy of Sciences, Xi'an 710119, China.

4. University of Chinese Academy of Sciences, Beijing 100049, China.

5. Nanophotonics and Metrology Laboratory, Swiss Federal Institute of Technology Lausanne, Lausanne 1015, Switzerland.

6. Institute for Advanced Study, The Hong Kong University of Science and Technology, Hong Kong SAR, 999077, China.

[†]These authors are contributed equally: T. Li, W. Gao, B. Fu

[*]Correspondence: xuxiaohao@opt.ac.cn; wangshuming@nju.edu.cn; yaobl@opt.ac.cn; phchan@ust.hk

# Abstract

Optical manipulation techniques offer exceptional contactless control but are fundamentally limited in their ability to perform parallel multitasking. To achieve high-density, versatile manipulation with subwavelength photonic devices, it is essential to sculpt light fields in multiple dimensions. Here, we overcome this challenge by introducing generalized optical meta-spanners (GOMSs) based on metasurfaces. Relying on complex-amplitude modulation, this platform generates lens-free, customizable optical fields that suppress diffractive losses. As a result, several advanced functionalities are simultaneously achieved, including longitudinally varying manipulation and in-plane spanner arrays, which outperforms the same operations realized by conventional donut-shaped orbital flows. Furthermore, the particle dynamics is reconfigurable simply by switching the input and output polarizations, facilitating robust multi-channel control. We experimentally validate the proposed approach by demonstrating single-particle dynamics and the parallel manipulation of particle ensembles, revealing exceptional stability for multitasking operations. These results demonstrate an ultracompact platform that is scalable to a larger number of optical spanners that operate simultaneously, advancing metadevices from wavefront sculptors to particle manipulators. We envision that the GOMS will catalyze innovations in cross-disciplinary fields such as targeted drug delivery and cell-level biomechanics.

# Introduction

The transfer of momentum from light to matter gives rise to optical forces, a phenomenon underlying to the principles of optical tweezers (OTs) [1-3]. OTs utilize the intensity gradient of a tightly focused Gaussian beam to generate a three-dimensional restoring potential, enabling the non-contact trapping and precise manipulation of micro- and nano-scale objects [1-6]. Beyond simple trapping, the exchange of linear momentum facilitates other intriguing effects such as particle pulling [7-9], optical binding [10-13], and laser cooling [14, 15]. The diversity of optical forces expands with more complex light fields. For instance, Laguerre-Gaussian beams carry orbital angular momentum (OAM), which provides a rotational degree of freedom (DoF) to optical manipulation [16-24]. These beams are characterized by an optical vortex—a helical phase front and a null intensity core—around which the Poynting vector spirals [25-27]. This beam profile exerts an optical torque or azimuthal force on particles [28], forming the basis of optical spanners (OSs, Fig. 1a) that are applied in optical sensing [29, 30] and microfluidics [31].

Precise optical manipulation with OSs requires subtle control over the phase and amplitude of the optical field. However, conventional devices for generating OAM face significant limitations. For example, spiral phase plates [32, 33] are restricted to generating conjugate topological charges, while spatial light modulators (SLMs) [34] introduce system complexity and are typically limited to a single hologram per configuration. On the other hand, metasurface have emerged as a miniaturized solution, enabling streamlined, multidimensional OAM generation [35]. One prominent example is an ultrathin J-plate that can integrate arbitrary OAM patterns into a single element [36, 37]. Further extension of the OAM control has been recently demonstrated with the total angular momentum plate (TAM) [38-40], This type of metasurface provides spatially varying topological charges along the optical field propagation direction, thereby unlocking a new degree of freedom for parallel, multitasking information processing [41-43].

Although metasurfaces have been found helpful in overcoming the issues of the bulk systems based on the discrete optics, their efficiency in advanced applications of OSs remain limited. This is because the primary function of the OS-oriented metasurfaces has long been altering only the force magnitude, but not controlling the particle trajectory [44, 45]. As a result, OS operations have appeared visually and physically monotonous. Only after the recent advent of generalized optical vortices (GOVs), this paradigm has been broken, enabling the shaping of OAM mode profiles into customized, non-circular geometries [46-49]. Unlike conventional OAM modes, which typically exhibit a ring-shaped intensity maximum whose radius is determined by the topological charge $\ell$, with larger $\ell$ values corresponding to intensity peaks farther from the center, the intensity distribution of a GOV varies with radial distance from the center. Theoretically, the topological charge of a GOV is a mixture of different $\ell$ values. Previous implementations of GOVs have relied on azimuthal phase modulation in Fourier space, requiring a lens for projection into real space. However, this approach inevitably introduces diffraction losses at the imaging center [48, 49], fundamentally limiting their potential as general optical spanners (GOSs). Similarly, spatial caustics fields, another method for engineering propagation-variant light fields, are also unsuitable as GOS candidates due to their uneven intensity distribution and lack of a phase gradient along closed paths [50].

To concurrently address the limitations of bulky systems with fixed patterns, and the growing demand for parallel multitasking micromanipulation, a portable GOS platform is required. As depicted in Fig. 1b, such a device would serve as an advanced successor to J-plates and TAM plates. It must encode OAM modes into arbitrary intensity contours, while preserving multifunctionality through controls such as incident-light switching and propagation-variation with mixtures of district $\ell$ values, thereby enabling complex and dynamic optical manipulation.

In this work, we demonstrate two classes of metasurface-based GOSs, each driving distinct rotational motions. The first class generates multiple GOVs along the longitudinal direction, while the second does this within the same transverse plane. Unlike prior GOV plates, our complex-amplitude metasurfaces transcend phase-only modulation in Fourier space by superimposing multiple Bessel rings to directly generate structured light fields without requiring external lenses, possessing phase-gradient forces to concurrently supply the intensity-gradient forces essential for trapping particles in polygonal trajectories during rotation. Furthermore, we employ circular polarization (CP) as the switching key, where two orthogonal CP channel pairs and one common CP channel provide the maximum efficiency across three independent channels. Both ultracompact GOSs enable on-chip particle delivery under configurable GOV profiles, showcasing customized micromanipulation capabilities. Crucially, this platform is highly scalable via arbitrary light-field engineering of the metasurface, the number and types of GOSs operating concurrently can be readily expanded beyond those demonstrated here.

**Results**

**Concept of generalized optical meta-spanner**

Fig. 1c presents a schematic of a generalized optical meta-spanner (GOMS), actuated by a propagation-varying generalized optical vortex (PVGOV). The system features two rotating wavefronts, represented by different colors along the optical axis, which enable particle delivery across varying longitudinal depths. Under a circularly polarized (CP) plane wave illumination, the device performs three distinct operational modes: when an incident left- or right-handed CP (LCP or RCP) light is filtered through an analyzer, the output exhibits reversed chirality (RCP or LCP), each carrying a unique GOV at different transverse planes. Additionally, filtering the same handedness of incident CP (LCP or RCP) yields a co-handed output channel, thereby generating an additional GOV mode. Together, these operations provide six distinct types of the GOVs. The rotational capability arises because the GOV possesses a helical phase $\varphi$, introduced via the OV flux relation $\boldsymbol{r} \times \boldsymbol{g}_\perp$ with $\boldsymbol{g} \propto I(r)\nabla\varphi$ [22], which provides both an intensity-gradient force and a phase-gradient force. The intensity-gradient force confines microscopic objects near the GOV, while the phase-gradient force drives their rotation around it. This fundamental GOV structure underpins the particle-rotation capability of the GOS, which can be selectively observed at distinct planes in free space.

**PVGOV generation.** Differing from conventional OT systems, GOMS directly projects the sculpted GOV field into real space without requiring extra lens-focusing, and therefore yields full complex-amplitude modulation. In practice, this operation is performed by two sequential steps: (1) acquisition of a phase-only GOV in the Fourier domain, and (2) superposition of the resulting Bessel rings to ensure intensity mapping. Specifically, consider a plane wave of uniform intensity $I_0$ that, upon passing through the Fourier-space modulator, acquires a transverse momentum shift $\Delta k$. The azimuthal angle of the resulting intensity profile relative to the optical axis in real space is $\theta$, while the azimuthal angle of this wave vector in Fourier space ($k$-space) is $\theta' = \theta + \frac{\pi}{2}$. The modulated intensity $I'_0$ in $k$-space and the original intensity $I_0$ in real space are related by the conservation of optical power $I'_0 \Delta k d(\Delta k) d\theta' = I_0 r dr d\theta$ in ray-optics approximation. When the beam is emitted along the optical axis, the cylindrical coordinates in $k$-space can be approximated as $(\Delta k, \theta', k\cos\theta_t) \approx (\Delta k, \theta', k)$. Consequently, for any closed contour in $k$-space, the resulting intensity distribution in radial and azimuthal coordinates $(k_\perp, \theta')$ satisfies $I_0'(k_\perp, \theta') \propto \frac{d\varphi(\theta)}{d\theta}$, which makes it possible to generate an arbitrary contour GOV. The detailed procedure for retrieving the exact phase profile is described in the Methods section.

Importantly, the optical force governing particle motion comprises both a phase-gradient force and an intensity-gradient force. At sharp turns or corners of the contour, the phase profile $\frac{d\varphi(\theta)}{d\theta}$ undergoes rapid changes, effectively resulting in a drastic phase change over an extremely short azimuthal range, and therefore leading to a localized Intensity gradient induced trapping force. As a result, the intensity-gradient force can dominate over the phase gradient force in these regions, trapping the particle and preventing continuous orbital motion. To avoid this and ensure smooth rotation, the phase gradient must be kept uniform along the entire closed path. Therefore, to drive a particle into continuous rotation with a GOV, it must interact with such a uniform phase profile. This is achieved through equal-arc-length sampling: one first selects the total number of sampling points, computes the total contour length via arc-length integration, and then assigns phases uniformly according to the arc length rather than the azimuthal angle [51].

After uniformly distributing the total arc length of the target contour among the chosen number of sampling points, we calculate the cumulative arc length from the initial sample to each subsequent point. By evaluating the difference between each actual sampled length and its corresponding target length, while locating the position in space where this difference vanishes, we recover the precise radial coordinate $k_\perp$ and azimuthal angle $\theta'$ for each point in the Fourier space, thus enforcing a uniform azimuthal phase gradient via $k_\perp(\theta') \propto \frac{d\varphi(\theta)}{d\theta}$. Next, we address the generation of two distinct phase patterns. The amplitude–modulated field $\Psi^{\alpha \to \beta}$ is designed to spatially deform the GOV from shape $\alpha$ to $\beta$ along optical axis, as schematically depicted in Fig. 1d. By applying a spatial Fourier transform to the field localized at $z = 0$ we obtain a longitudinal amplitude $\tilde{\mathcal{A}}$ confined within the range $[L_1, L_2]$,

Because $\tilde{\mathcal{A}}$ is centrosymmetric and azimuthally invariant (see Method), variations in the azimuthal phase do not alter the longitudinal intensity profile within that interval. Therefore, to generate a GOV field of the desired shape over $[L_1, L_2]$, we simply impose a designed phase $\varphi_\alpha(\theta)$ onto the component $\mathcal{A}(L_1, L_2, k_\perp)$. Similarly, to create the second GOV segment for OS within $[L_2, L_3]$, we apply the phase $\varphi_\beta(\theta)$ to $\mathcal{A}(L_2, L_3, k_\perp)$. The field at any longitudinal position $z$ can then be reconstructed following the procedures described in the Methods section. Figure. 1d shows a ray-tracing diagram together with simulated transverse intensity profiles along the propagation axis, with ray colors indicating the local phase variation at each pixel.

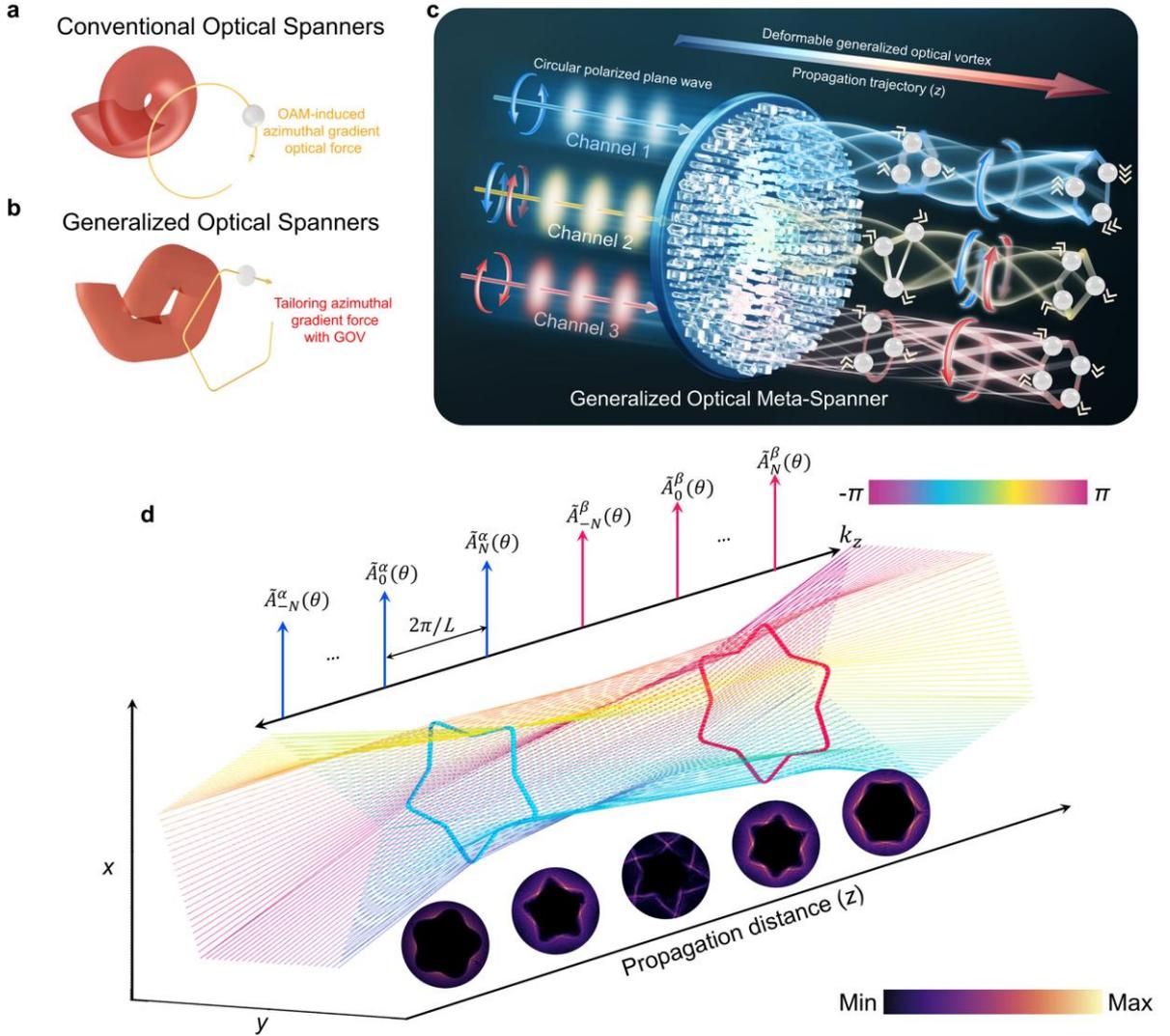

**Fig. 1**. Concept of GOMS. a. Conventional OS, which induces an azimuthal phase-gradient force via the OAM components of higher-order Laguerre–Gaussian beams. b. GOS of arbitrary polygonal shape defined by a closed contour in the transverse plane. c. Schematic representation of the GOMS, which combines portability with multitasking. By utilizing three CP channels, it can generate PVGOV fields to drive particles into multifaceted motion. d. Principle of propagation-varying GOV design. Each optical field pattern Ψ comprises 2$N$+1 Bessel functions as an equally spaced comb in $k$ space. The ray tracings color encodes the local phase, while corresponding simulated transverse intensity profiles at different propagation distances are depicted below. By precisely weighting these profiles, the on-axis intensity distribution along $z$-direction follows a tailored intensity $F(z)$ resulting in a contour transition from a pentagon to a hexagon.

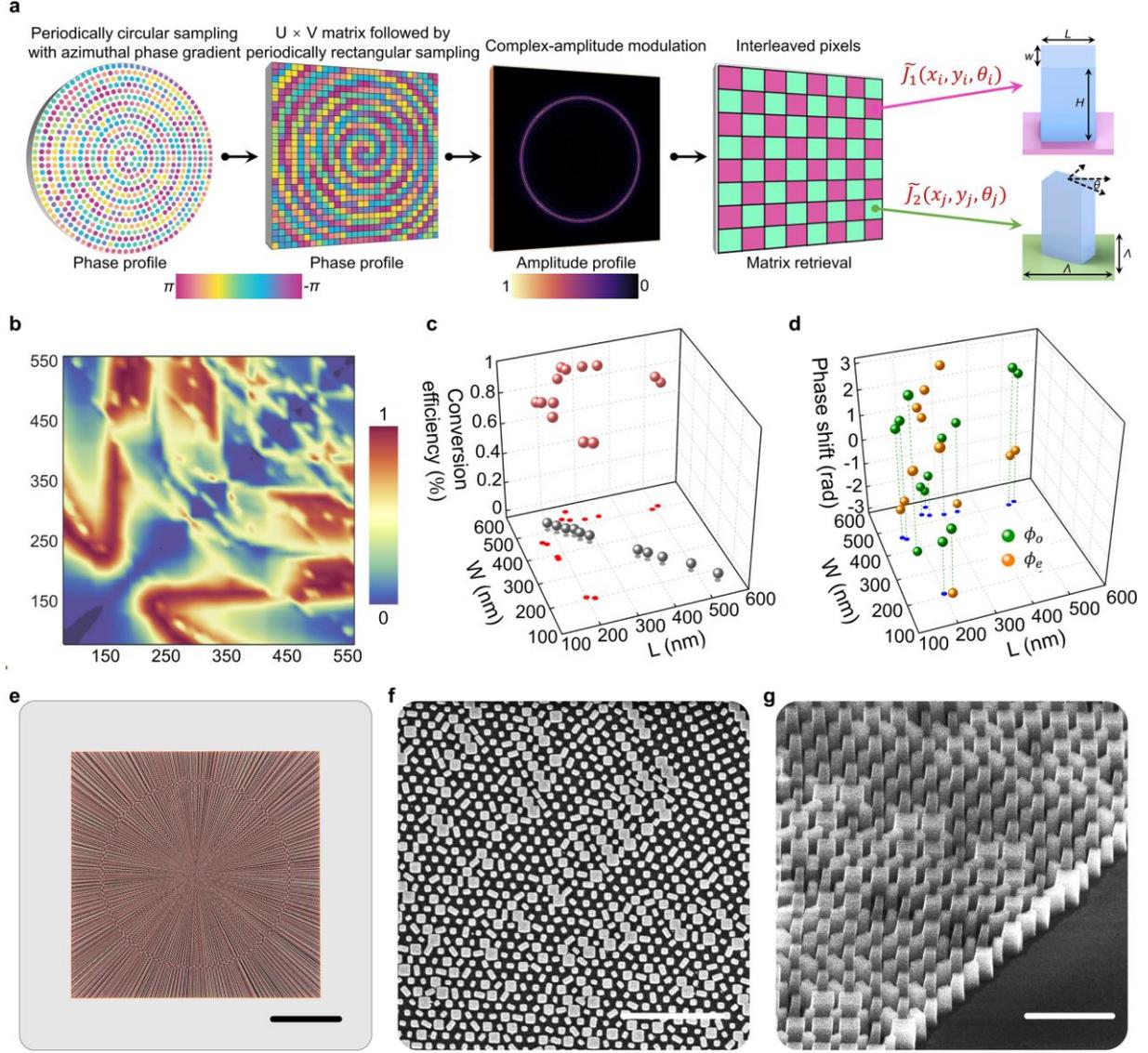

**Fig. 2**. Design and fabrication of the multitasking metasurface. a. Flowchart of the metasurface encoding process: sampling each pixel in polar coordinates and remapping to a Cartesian matrix; applying amplitude modulation; and retrieving the corresponding nanopillar geometry for pattern filling. Nanopillars for channels $\tilde{J}_1$ and $\tilde{J}_2$ are interleaved spatially. b. Simulated CP conversion efficiency of the designed nanopillars. c. Measured conversion efficiencies of the selected nanopillars for the cross-CP channels (red) and the co-CP channel (gray). d. Phase retardance profiles $\phi_o$ and $\phi_e$ corresponding to the data shown in (c). e. Optical micrograph of the fabricated metasurface, scale bar, 250 μm. f, g. Top-view and side-view SEM images of the metasurface, scale bar, 5 μm.

**Multitasking metasurface implementation.** Based on the above conceptual procedure for generating a PVGOV, we implement a Jones-matrix–engineered metasurface to perform the required GOV transformation. As illustrated in Fig. 2a, the radial coordinate and azimuthal angle that define the azimuthal phase gradient are first mapped into a U×V matrix for sampling. This matrix is then combined with the complex-amplitude modulation to yield the total phase profile. To integrate multiple, distinct phase patterns onto an individual metasurface, each nanopillar is designed to exhibit a strong refractive-index contrast across four CP channels; this behavior is captured by the Jones-matrix formalism $J(x,y) = R(\theta) \begin{bmatrix} T_o e^{i\phi_o(x,y)} & 0 \\ 0 & T_e e^{i\phi_e(x,y)} \end{bmatrix} R(-\theta)$ [51-53], where $\phi_o$ and $\phi_e$ denote the pillar's phase shifts for the ordinary and extraordinary axes, respectively, and $R(\theta)$ is the in-plane rotation matrix by the pillar's orientation angle $\theta$. Due to the unitarity of the Jones matrix, the co-CP channels simply replicate each other without polarization conversion, while the cross-CP channels exploit the birefringent difference between the nanopillar's ordinary and extraordinary axes. The metasurface is assumed to perform as a pure phase shaper with equal transmittance ($T_o = T_e = T$) for both axes to achieve maximum CP conversion efficiency, therefore the Jones matrix defined in the CP basis $[e_L \ e_R]^T$ can be simplified to a purely diagonal form $J(x,y) = T e^{i\phi_r(x,y)} \cdot \begin{bmatrix} \cos\left(\frac{\delta}{2}\right) & i\sin\left(\frac{\delta}{2}\right) e^{-i2\theta} \\ i\sin\left(\frac{\delta}{2}\right) e^{i2\theta} & \cos\left(\frac{\delta}{2}\right) \end{bmatrix}$. Here, the phase difference $\delta$ governs the polarization state of both the incident and transmitted beams (see Supporting Information Note 1 for a full derivation). To implement three independent GOV channels simultaneously, we decompose the overall Jones matrix as $J = \tilde{J}_1 + \tilde{J}_2$, where $\tilde{J}_1(x_i, y_i, \theta_i)$ and $\tilde{J}_2(x_j, y_j, \theta_j)$ correspond to two different $\delta$, These sub-matrices are then interleaved across the metasurface as shown at the end of the encoding flowchart. As the unit cell, we employ amorphous silicon ($a$-Si) nanopillars with specific height ($H$), length($L$), width ($W$), and in-plane rotation angle ($\theta$) grown on fused-quartz substrates; the pillars stand freely in air on their top faces. Their refractive index dispersion is provided in Supporting Information Note 2. For each nanopillar and working wavelength $\lambda$ = 1064 nm, we numerically obtain a phase library, which is depicted in Fig. 2a by a specific color. Fig. 2b shows the simulated CP conversion efficiencies across our phase-library of nanopillars. From this library, we select two sets of nanopillars optimized for the cross-CP and co-CP channels, respectively. Their geometric parameters, and the corresponding ordinary and extraordinary phase shift are listed in Fig. 2c and 2d. The detailed phase diagram related to geometric dimension is shown in Supporting Information Note 3. Following the above encoding protocol, we map three distinct PVGOV phase profiles onto a single, monolithic metasurface and fabricate the resulting GOMS using standard semiconductor microfabrication techniques. An optical-microscopy image of the fabricated device is shown in Fig. 2e, and higher-resolution top- and side-view SEM images are depicted in Figs. 2f and 2g, respectively. The detailed fabrication procedure is provided in the Methods section.

**Light field characterization.** We first experimentally study the GOMS fields using a standard 4$f$ imaging setup, with each channel incorporating polarization preparation and analysis stages to isolate three independent GOV outputs (see Supporting Information Note 4 for the full optical schematic and alignment details). Fig. 3 displays the measured transverse intensity distributions, pseudo-colored by channel.

Fig. 3a shows the triangular–to–quadrilateral GOV evolution with its total phase profile $\varphi_1$ under LCP illumination and RCP analysis. The bottom part of this figure also depicts the theoretical phase profile and amplitude of the GOV in the Fourier space. The outgoing beam initially forms a defocused speckle near the sample plane. Upon propagation, this speckle converges into a well-defined triangular GOV at $z$ = 1.2 mm, whose clearest transverse plane is chosen as the first GOS. As the beam continues to propagate, the intensity reverts to a speckle pattern, evolves into a four-lobed distribution, and finally converges into a well-formed quadrilateral GOV at $z$ = 3.6 mm. We extract the uniform-intensity slice as the second GOS target. Fig. 3b presents the pentagon–to–hexagon transformation in the co-CP channel with LCP or RCP input and LCP or RCP output with their total phase $\varphi_2$ and specific Fourier-space information. In this case, the field first focuses into a pentagonal GOV, then returns to a speckle, and subsequently converges into a hexagonal GOV. Fig. 3c shows the donut-shape OAM (RCP input and LCP output) evolving into a heptagonal GOV upon propagation.

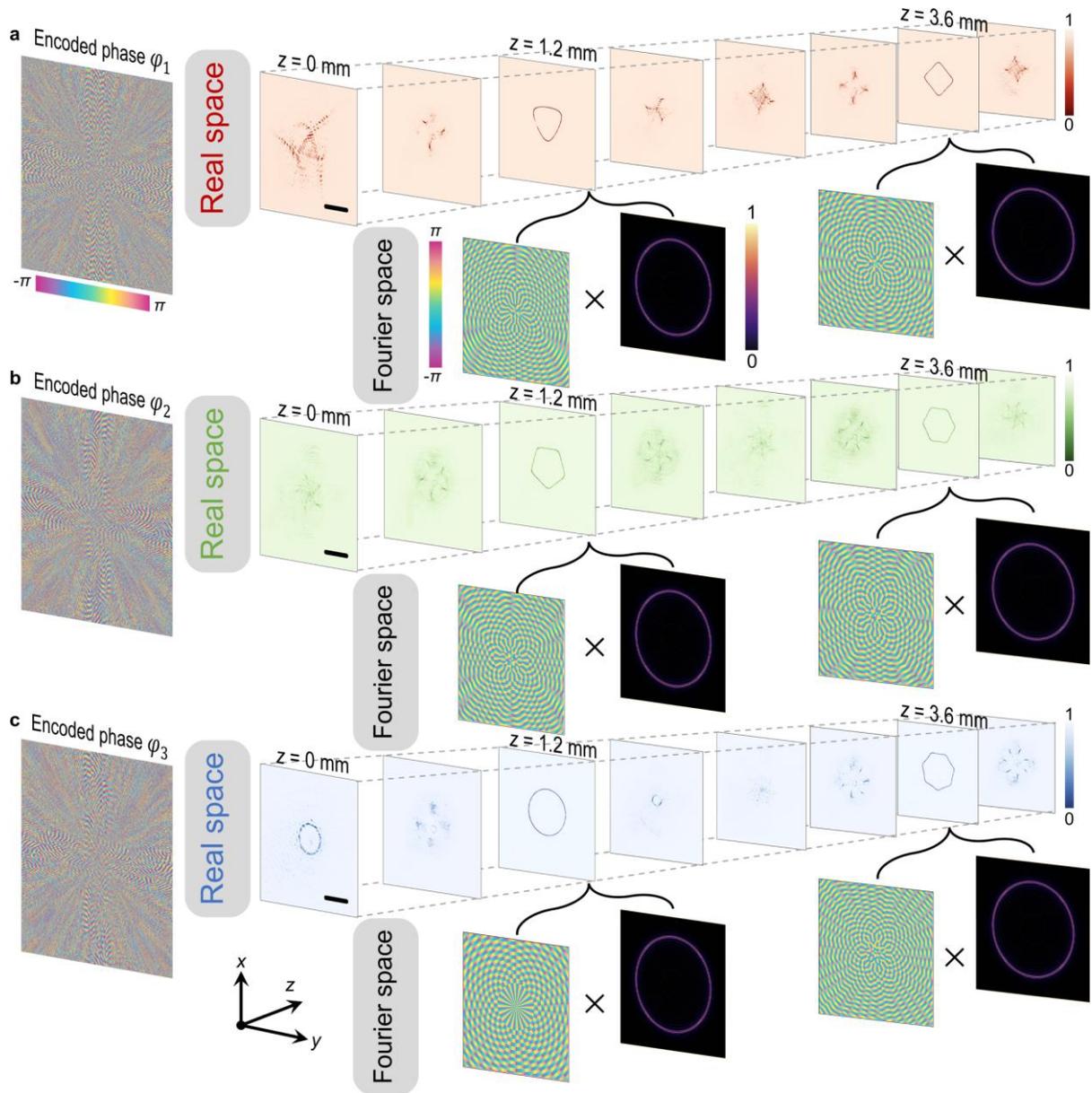

**Fig. 3**. PVGOV characterization of metasurface. a-c. The experimental measurements of the PVGOVs, which evolves from triangular to quadrilateral, pentagonal to hexagonal, and donut-shaped to heptagonal GOVs under three distinct CP channels, respectively, along with the total phase encoded in each channel and the corresponding ideal phase and amplitude profiles in Fourier space. Scale bar: 10 μm.

**Micromanipulation demonstration for GOMS.** We next demonstrate the micromanipulation capability of our system by using three distinct GOV shapes and gold particles with diameters ranging from 0.8 to 3 µm. The optical setup used for these experiments is schematically illustrated in Fig. 4a, and its detailed description can be found in the Methods section. Fig. 4b depicts the microfluidic trapping chamber, which consist of a pair of microscope slide and coverslip separated by the gap of $h_d$ = 140 µm and filled with the suspension of gold particles (refractive index $n_{gold}$ = 0.26 + 6.97$i$) in water ($n_s$ = 1.33). The forces acting on each particle are the scattering force $F_{sca}$ due to radiation pressure, the intensity-gradient force $F_{grad}$, the gravitational (G) and buoyancy, the drag force and the reaction force $F_{cover}$ from the coverslip, which together establish mechanical equilibrium.

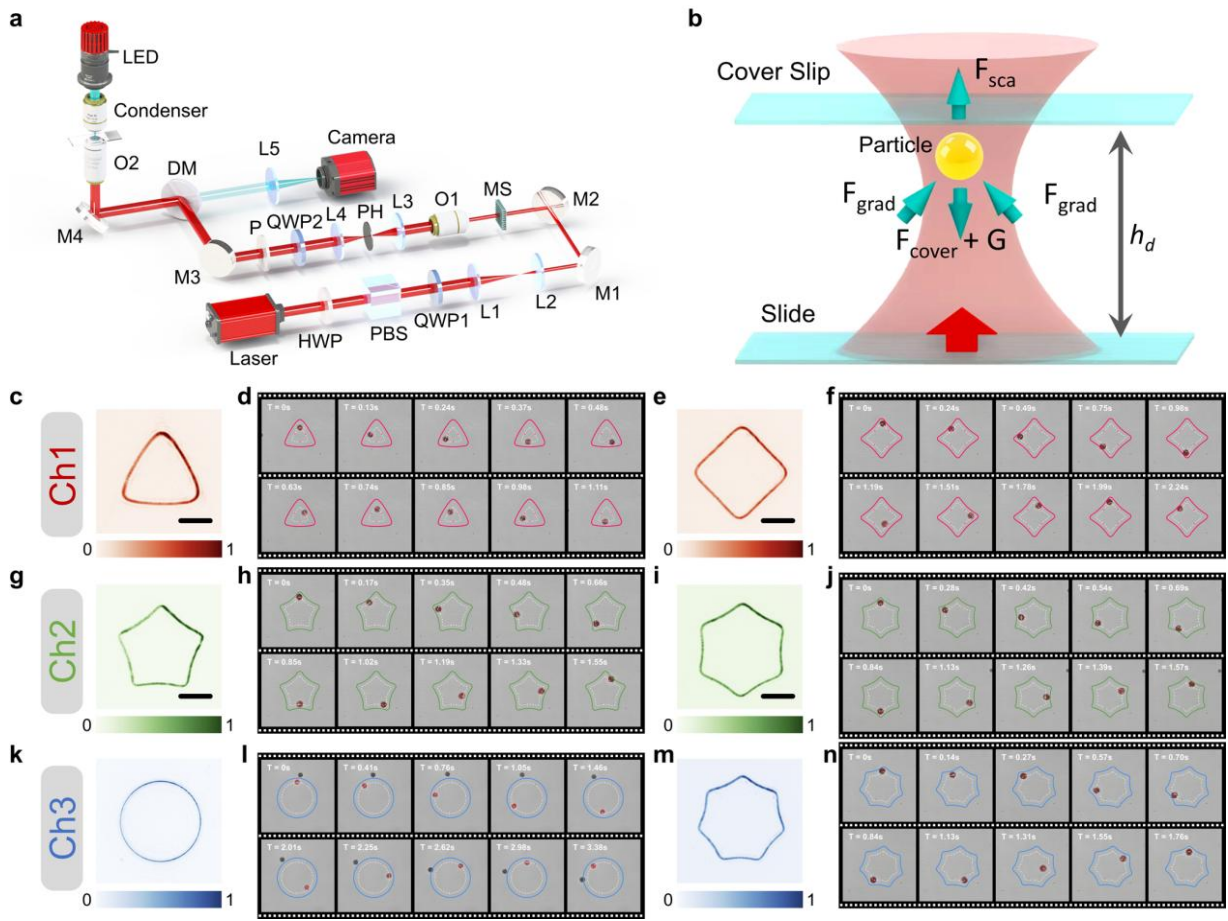

**Fig. 4**. Micromanipulation demonstration. a. Schematic of the experimental setup. b. Detailed view of the trapping chamber. c-n. Time-lapse micrographs of gold particles being manipulated along the GOV trajectories (dashed arrows) by the targeted GOV contours (pseudocolored lines). Scale bar, 5 µm.

Figs. 4c–f show the intensity distributions and corresponding micromanipulation snapshots for channel 1. Under LCP illumination and RCP analysis, the gold particle is confined to the vicinity of the triangular GOV pattern (Fig. 4c) at a distance of about $z$ = 1.2 mm and experiences a counter-clockwise (CCW) orbital motion along the triangular contour. The time-resolved trajectories of this movement are depicted in Fig. 4d, demonstrating stable particle trapping inside the triangular GOV and continuous rotation. By spatially moving the metasurface to reposition the field on the sample plane—from triangle to quadrilateral (Fig. 4e)—the particle similarly undergoes uniform, stable CCW rotation within the primary lobe, as shown in Fig. 4f. Rotating HWP2 by $\pi/2$ switches to channel 2. This new channel produces the pentagonal GOV at the imaging plane (Fig. 4g). The time-lapse sequence depicted in Fig. 4h captures the particle confinement within the pentagon and movement along its edges. By translating the metasurface closer to the source reveals the hexagonal GOV (Fig. 4i), and the particle dynamics is recorded in the time-lapse series of Fig. 4j.

For the third GOV provided by the channel 3, we switch the input RCP and LCP analysis, which produces a conventional donut-shaped OAM beam at one focal plane and a heptagonal GOV at another (Fig. 4k, m). In both cases, the particle rotates CCW along the inner contour, as seen from the time-lapse snapshots and their timestamps in Fig. 4l and 4n. Additionally, the particle can also be trapped and continuously rotated outside these spanner contours (see Supporting Information Note 5). The Supporting Information Note 6 also details our numerical calculations of the lateral optical forces exerted by each GOV on the gold particle.

**Necklace rotation via metasurface-enabled GOSA**. Beyond longitudinal multiplexing, our metasurface can simultaneously encode multiple GOV fields in the same plane to form a transverse, multiplexed generalized optical spanner array (GOSA). As illustrated in Fig. 5a, the idea is to encode each GOV phase profile independently and superimpose it with a unique deflection phase $\varphi_j^d = kp\sin\chi_j$, where $p$ is the direction of light field and $\chi$ is the deflect angle. This produces an in-plane GOSA capable of parallel, multitarget micromanipulation. The composite phase distribution loaded onto the metasurface is $\varphi_{GOVA} = \sum_{j=1}^{N} \arg\left[e^{i(\varphi_j^{GOV} + \varphi_j^d)}\right]$. The overall phase-encoding workflow is depicted in Fig. 5a. For each individual GOV, we superimpose a pair of deflection angles $(\chi_j^x, \chi_j^y)$ onto the off-axis coordinates $(x, y)$, the sum of these deflections and the basic GOV phase yield the total phase profile. The three encoded total phases are shown in Supporting information Note 7. Figs. 5b–d show the measured focal-plane intensity patterns for these three GOSA channels. Each GOVA is addressable independently, and switching channels reshapes the field distribution. To demonstrate parallel manipulation, we simultaneously trap four arrays of 3 μm-diameter $SiO_2$ beads at the focal plane and recorded their motion (Figs. 5e–g). As can be seen from the figures, the particles in the trapping regions are arranged into chains resembling necklace shapes. Using blue-dye–labeled beads as fiducials, we observe that the upper-left and lower-right necklaces, which have opposite topological charges, rotate CCW, whereas the other two necklaces (tracked via red-dye beads) rotate clockwise (CW). Crucially, all four necklaces move without mutual interference, confirming the possibility to simultaneously and independently control the particle orbital movement with the metasurface-enabled GOSA. This parallel processing requires no additional optical elements, offering a new avenue for multitasking manipulation.

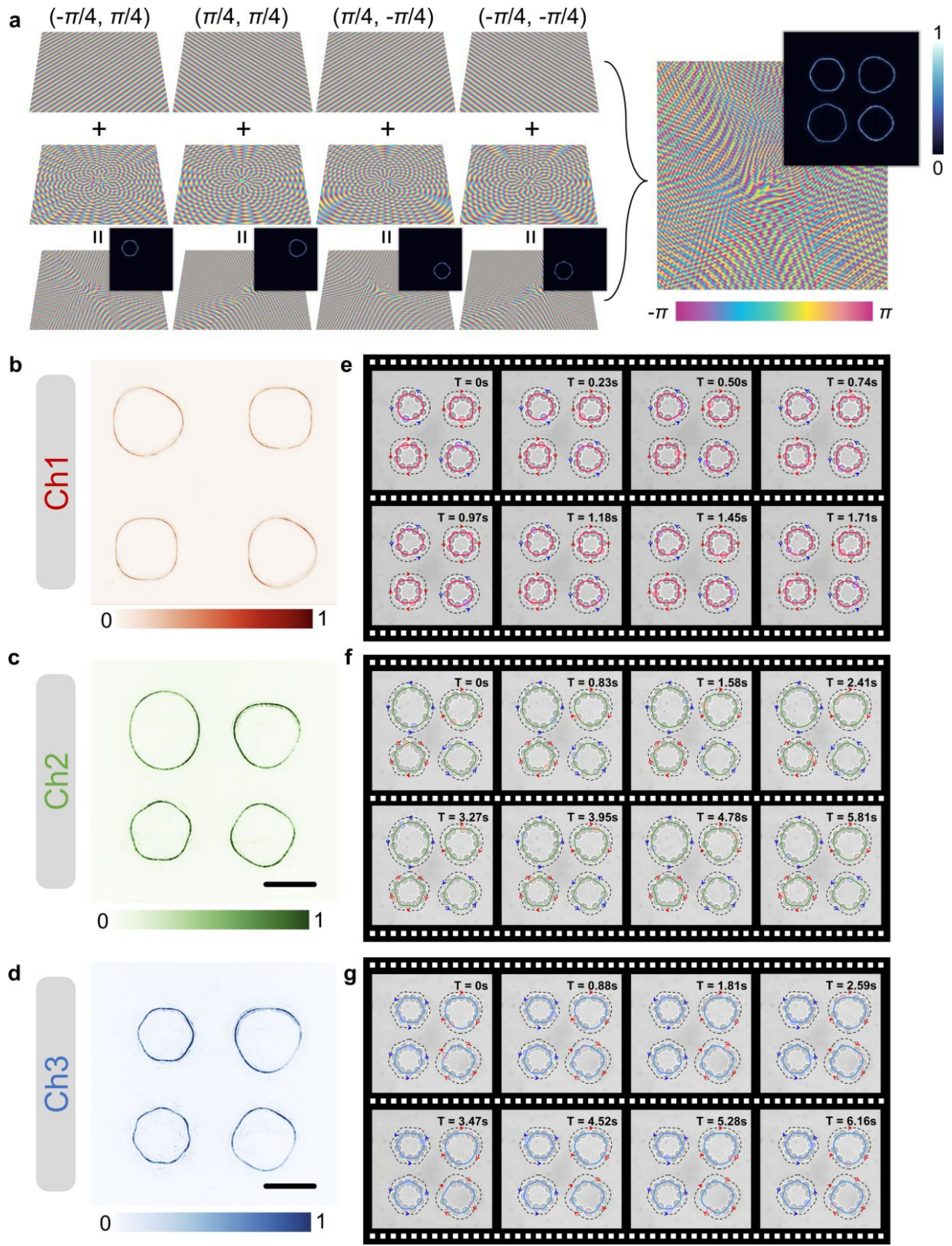

**Fig. 5**. Metasurface-based GOSA. a. The design flow and phase distribution of GOSA. Each phase profile is obtained by independently calculating and linearly superimposing phase gradient to the four directions respectively, and finally combining them with the 3-channel metasurface. b-d. The experimental results of the light field distribution

corresponding to the total phase in (a). e-g. Time-lapse imaging of multi-$SiO_2$ necklace manipulation using three channels of the metasurface. These time-lapse snapshots were obtained using the in-plane multitasking GOSA depicted in (b-d), respectively. In each case, the upper-left and lower-right necklaces exhibit CCW rotation, while the upper-right and lower-left necklaces demonstrate CW rotation.

**Discussion**

In summary, we report a metasurface-based generalized optical spanner (GOS) that combines propagation-varying generalized optical vortices (GOVs) with full complex-amplitude control, overcoming the field-shaping and single-function limitations of conventional optical micromanipulation. By leveraging both the intrinsically polarized degrees of freedom and the propagation axis, our platform realizes on-axis varying GOV and simultaneously generates multiple GOV array traps in the focal plane for differentiated rotational tasks. We introduce a unified complex-amplitude design paradigm that synergistically optimizes the azimuthal phase-gradient torque and self-focusing intensity-gradient forces, enabling arbitrary-shaped optical spanners with both trapping and rotation capabilities, thereby resolving the inherent trade-off between torque diversity and trapping stability in phase-only devices. An orthogonal circular-polarization channel-switching scheme provides ultra-efficient, multi-task, reconfigurable control. This ultracompact platform not only obviates the complexity and limited functionality of bulky system and integrates with optical fibers or waveguides [54-56], but also extends the scope of optical micromanipulation through space-programmable GOV landscapes.

The broader impact of our GOMS lies in other disciplines, offering distinct capabilities for fundamental science and advanced manufacturing. For instance, it provides a unique testbed for investigating fluctuation theorems through simulated entropy variation dictated by trajectory winding numbers [57], and for conceiving continuous-time crystals with customized looped lattice through force field [58]. Practically, its parallel manipulation scheme enables novel microfabrication paradigms, such as using multiple trapped particles that are self-assembled and dynamically reconfigurable into micro-nozzles for mask-free polymer extrusion [59, 60].

**Methods**

**Principle of PVGOV generation**

According to the generalized Snell's law, the abrupt phase change in the azimuthal direction at the air-interface under normal incidence must satisfy the condition $n_t \sin(\theta_t) = \frac{1}{k_0 r} \cdot \frac{d\varphi(\theta)}{d\theta}$, where $n_t$ is the refractive index of the medium after transmission. In Fourier space (*k*-space), where a position is described by the transverse wavevector $k_\perp$, this phase gradient corresponds to a local change in the transverse wave vector. In cylindrical coordinates, the phase gradient is described as $\nabla_\perp \varphi = \frac{1}{r} \frac{d\varphi}{d\theta} \hat{e}_\theta$, and its magnitude defines the additional momentum $\Delta k$ as $|\Delta k| = |\nabla_\perp \varphi| = k_\perp \sin\theta_t \approx n_t k_0 \sin\theta_t$, where the azimuthal angle of this wave vector in *k*-space is $\theta' = \theta + \frac{\pi}{2}$ with the unit vector is $\hat{e}_{\theta'}$. When a plane wave with intensity $I_0$ impinges on this phase-discontinuous interface, the intensity distribution in *k*-space $I'_0$ can be related to the real-space intensity through power conservation. For a differential area on the metasurface $rdrd\theta$, the corresponding power in *k*-space is $I'_0 \Delta k d(\Delta k) d\theta' = I_0 r dr d\theta$. Using the relation $\Delta k = \frac{d\varphi(\theta)}{r d\theta}$ (for a fixed $\theta$, treating $d\varphi/d\theta$ as constant), the differential $dr$ can be expressed as $dr = -\frac{1}{\Delta k^2} \frac{d\varphi}{d\theta} d(\Delta k)$. Substituting and solving for $I'_0$ gives the intensity profile in *k*-space [48]:

$$I'_0 = \begin{cases} \frac{I_0}{\Delta k^4} \left(\frac{d\varphi(\theta)}{d\theta}\right)^2, & \Delta k \geq \Delta k_{\min}(\theta) \\ 0, & \Delta k < \Delta k_{\min}(\theta) \end{cases} \quad (1)$$

Here, $\Delta k_{min}(\theta)$ represents the minimum change in wave vector corresponding to the scattered light at angle $\theta$, which occurs $\Delta k_{min}(\theta) = \frac{d\varphi(\theta)}{r_m d\theta}$ at the outermost radius of the metasurface $r_m$. This equation reveals that under the ray optics approximation, the intensity peaks at $\Delta k = \Delta k_{min}$ and decays with the fourth power of $\Delta k$. Consequently, the most intense contour in *k*-space defined by $\Delta k = \Delta k_{min}$ has an intensity profile proportional to the phase gradient. Using the coordinate correspondence $\theta' = \theta + \frac{\pi}{2}$, we find the modulated intensity $k_\perp(\theta') \propto \frac{d\varphi(\theta)}{d\theta}$. This simple relation provides a direct design rule for generating arbitrary closed-loop intensity profiles, one can inversely engineer GOV with the azimuthal phase profile $\varphi(\theta)$ accordingly.

For longitudinally varying GOVs, we employ the frozen wave principle by integrating multiple phase profiles. Suppose the target axial intensity distribution is $F(z)$, over $z \in [0, L]$. This can be realized as superposition of $2N+1$ Bessel beams [38]:

$$\Psi(r, z) = e^{iQz} \sum_{n=-N}^{N} \tilde{A}_n J_0(k_{rn} r) e^{2\pi n z i / L} \qquad (2)$$

where $Q$ represents the localized parameter of transverse intensity, $N$ is an integer satisfying $N \ll (\frac{\omega}{c} - Q) L / 2\pi$, $k_{rn}$ denotes the transverse wavevector for the $n$-th mode with $k_{rn}^2 = \frac{\omega^2}{c^2} - (Q + \frac{2\pi n}{L})^2$, $J_0$ is the zeroth-order Bessel function, and the coefficients $A_n$ is given by $A_n = \frac{1}{L}\int_0^L F_z(z) e^{-2\pi n z i / L} dz$. The function $F_z(z)$ over the distance $L$ is the target $z$-dependent intensity chosen by design. Based on this approach, it is theoretically possible to customize a segment of quasi-constant intensity waves at any distance. Therefore, for a wave propagating within the range $z \in [L_1, L_2]$, i.e., $F(z) = 1$ in this interval, the field distribution becomes:

$$\Psi(r, z) = e^{iQz} \sum_{n=-N}^{N} \frac{i(e^{-2\pi n L_1 i / L} - e^{-2\pi n L_2 i / L})}{2n\pi} J_0(k_{rn} r) \qquad (4)$$

By applying the spatial Fourier transform, the spatial frequency spectrum (angular spectrum) of the optical field at $z = 0$, localized longitudinally within the range $[L_1, L_2]$, can be expressed as:

$$\tilde{\mathcal{A}}(L_1, L_2, k_\perp) = \frac{i}{\pi} \sum_{n=-N}^{N} \frac{i(e^{-2\pi n L_1 i / L} - e^{-2\pi n L_2 i / L})}{2n\pi} \begin{cases} \frac{1}{\sqrt{k_{rn}^2 - k_\perp^2}}, & k_\perp < k_{rn} \\ 0, & others \end{cases} \qquad (5)$$

This frequency spectrum distribution is centrally symmetric and azimuthally insensitive, with the intensity maximum located at $k_\perp \approx k_{rn}$. Therefore, azimuthal phase profile has no influence on longitudinal intensity distribution of the optical field. Suppose we generate a GOV field with a specific shape within the interval $[L_1, L_2]$ under a specific channel within metasurface, we simply need to apply a designed phase $\varphi_\alpha(\theta)$ to $\mathcal{A}(L_1, L_2, k_\perp)$ in the spectral space. Similarly, to generate another GOV field with a specific shape in the interval $[L_2, L_3]$, we apply the phase $\varphi_\beta(\theta)$ to $\mathcal{A}(L_2, L_3, k_\perp)$. In summary, the total spectrum for a longitudinally varying GOV transitioning from $\varphi_\alpha(\theta)$ to $\varphi_\beta(\theta)$ is $\mathcal{A}_{total}(\theta, k_\perp) = \mathcal{A}(L_1, L_2, k_\perp)e^{i\varphi_\alpha(\theta)} + \mathcal{A}(L_1, L_2, k_\perp)e^{i\varphi_\beta(\theta)}$. Thus, the GOV at any longitudinal position $z$ for one channel can be obtained through the inverse Fourier transform:

$$\Psi_{total}(r, \theta, z) = \frac{1}{(2\pi)^2} \iint \mathcal{A}_{total}(\theta, k_\perp) e^{i[k_\perp r\cos(\theta - \theta') - k_z z]} k_\perp dk_\perp d\theta' \qquad (6)$$

**Numerical simulation**

Our metasurface sample was designed by positioning α-Si nanopillars on a fused-silica substrate. The modulation phase generated was associated with the structural dimensions of the nanopillar. To find the optimal parameters conducive to the intended phase modulation, arrays of nanopillars with distinct geometric configurations were modeled using the Finite-Difference Time-Domain (FDTD) method. These nanopillars were organized in a square lattice having a lattice constant of $\Lambda$=600 nm and a uniform height of $H$=800 nm (Fig. 2a). The refractive indices of α-Si utilized were within the near-infrared wavelength range, determined from ellipsometry assessments of an α-Si thin film. The boundary conditions were designated as periodic in the in-plane directions to model the periodic nanopillar arrays, while for transmission situations, the boundary conditions were identified as perfect match layers in the light incidence direction. A plane wave with a wavelength $\lambda = 1064$ nm was employed to discern the transmission coefficients and phase shifts along the *x*- and *y*-axes, corresponding to the lateral and vertical dimensions of nanopillars.

**Sample fabrication**

The silica-based wafer underwent ultrasonic cleansing using acetone, alcohol, and deionized water. This was succeeded by the application of an 800 nm layer of amorphous silicon (α-Si) onto the pristine silica wafer via Plasma-Enhanced Chemical Vapor Deposition (Oxford PlasmaPro 100 PECVD). Afterward, the silicon surface received a spin-coated layer of conductive, positive electron-beam resist (ZEP520A, Zeon) approximately 200 nm thick. The intended metasurface designs were then etched onto the CSAR-6200 resist using an Electron Beam Lithography (EBL) system, specifically the Elionix ELS-F125-G8. Post-exposure, the patterns were developed in a photoresist developer, transferring them onto a chromium (Cr) mask. Subsequent steps involved depositing a patterned chromium (Cr) layer onto the sample via Electron Beam Evaporation (EBE), creating a Cr mask on the thin film for further etching processes. The etching was performed using an Inductively Coupled Plasma Reactive Ion Etcher (ICP-RIE) with a mixture of $O_2$ and $CHF_3$ gases. Finally, the remaining Cr mask was removed using a cerium ammonium nitrate solution, completing the fabrication of our metasurface. The visualized workflow refers in Supporting Information Note 8.

**Micromanipulation experiment.**

In this experiment, a continuous-wave fiber laser with a wavelength of 1064 nm (VFLS-1064-B-SF-HP, Connet Laser Technology Co., Ltd., China; rated power: 5 W) was employed as the light source. The laser beam was first modulated into horizontally linear-polarized light using a combination of a half-wave plate (HWP) and a polarizing beam splitter (PBS). It was then converted into circularly polarized (CP) light by passing through a quarter-wave plate (QWP1), with the handedness of the incident CP flexibly controlled by adjusting the orientation of QWP1. Subsequently, the beam passed through a collimating and beam-reducing system composed of lenses L1 and L2, was reflected by mirrors M1 and M2, and then compressed to precisely cover the metasurface (MS) mounted on a manual translation stage. The transmitted light modulated by the metasurface was collected by an objective lens O1 (40×, NA 0.95, CFI Plan Apo, Nikon Inc., Japan) and further processed by a spatial filtering system consisting of lenses L3 and L4 and a pinhole (PH). A complex spectrum was formed at the Fourier plane of L3, effectively filtering out higher-order diffraction components in k-space. Afterwards, polarization filtering was implemented using a combination of a quarter-wave plate (QWP2) and a linear polarizer (P) to extract the cross-polarized component of the transmitted field. This component was reflected by a mirror M3 toward a dichroic mirror (DM, DMR-1000SP, 1000 nm, LBTEK, China), where the infrared portion was reflected by mirror M4 into the back aperture of an oil-immersion objective lens O2 (100×, NA = 1.4, CFI Plan Apo, Nikon Inc., Japan), forming the desired optical trap at the focal plane. In the imaging part, a condenser lens focused LED illumination onto the sample chamber, while imaging of the sample was accomplished by O2 and L5. The entire manipulation process was monitored and recorded in real time by a CCD camera (Point Grey GS3U3-41C6M-C, FLIR Systems Inc., USA; 2048 × 2048 pixels, pixel size 5.5 μm, frame rate 90 fps).

**Acknowledgement**

The authors thank Sergejs Boroviks (EPFL) for his valuable suggestions on this work.

**Funding**

This work was supported by the National Program on Key Basic Research Project of China (2022YFA1404300), the National Natural Science Foundation of China (12325411, 62288101 and 11774162), the Open Research Fund of the State Key Laboratory of Transient Optics and Photonics, Chinese Academy of Sciences (SKLST202218), the Fundamental Research Funds for the Central Universities (020414380175), the Natural Science Foundation of Jiangsu Province (BK20233001) and the Jiangsu Provincial Key Research and Development Program (BG2024029). Work done in Hong Kong is supported by RGC Hong Kong (CRS_HKUST601/23 and AoE/P-502/20).

**Author contributions**

T.L. conceived the idea. T.L., W.G., and B.F. performed the experiments. T.L. and W.G. developed the theoretical framework and conducted the analysis. B.F. carried out the sample fabrication and characterization. The manuscript was written by T.L., S.Z., X.X., B.Y. and C.T.C. T.S., Y.F., J.K.N., S.Y., Z.W., O.J.F.M participated in scientific discussions and provided feedback on the manuscript. S.W., S.Z. and C.T.C. supervised the project.